\RequirePackage{fix-cm}
\documentclass[smallextended]{svjour3}       
\smartqed  
\usepackage{graphicx}
\usepackage{braket}
\begin{document}

\title{Bouncing and coasting universe with exact quantum-classical correspondence
}


\author{Moncy Vilavinal John    
}


\institute{Moncy Vilavinal John \\
              Department of Physics, St. Thomas College \\
              Kozhencherry, Kerala India \\
              Tel.: 91-9447059964\\
                         \email{moncyjohn@yahoo.co.uk}             \\
}

\date{Received: date / Accepted: date}

\maketitle

\begin{abstract}
When the scale factor  of expansion of the universe is written  as $ a(t)\equiv Ae^{\alpha(t)}$, with  $A$ as some real constant and $\alpha(t)$ a real function, the  gravitational action $I_G$  appears in the same form as  the matter action $I_M$ for a homogeneous and isotropic scalar field with a specific scale factor-dependent potential.     We observe that  by making analytic continuation of  the Lagrangian function in this $I_G$  to the complex plane of $\alpha$,  one can obtain terms corresponding to both parts of a total action  that  describes a viable cosmological model.  The solution gives a  bouncing and coasting universe, which first contracts and then expands linearly, with a smooth bounce in between them. Such a bounce, called a Tolman wormhole, is due to a Casimir-like negative energy that appears naturally in the solution. In a novel approach to quantum cosmology, we perform  canonical quantization of the above model using the operator method and show that it can  circumvent the operator ordering ambiguity in the conventional formalism. It leads to a quantum wave equation for the universe, solving which we get the interesting result that the universe is in a ground state with nonzero energy. This solution is different from  the Wheeler-DeWitt wave function and    possesses exact quantum-classical correspondence during all epochs.

\end{abstract}

\section{Introduction}
 The gravitational action $I_G$ in general relativity for a spacetime described by the Robertson-Walker metric can be written in the same form as  the matter action $I_M$  corresponding to a homogeneous and isotropic scalar field with a specific scale factor-dependent potential.  In this case  $I_M$, when taken alone,  is capable of giving the quantum mechanical wave equation for the scalar field. It is traditionally viewed that general relativity and quantum theory are disparate physical theories. A new paradigm that emerges in modern physics is that contrary to this viewpoint, gravity and quantum mechanics are closely related theories \cite{susskind}.   The above case of an equivalence between $I_G$ and $I_M$ may also  be considered as suggestive of some close relationship between gravity and quantum mechanics.

 The first part of  the present paper   shows that by a complex extension of  the gravitational action $I_G$, one can reach a  bouncing, eternal coasting cosmological model.  To show this, we first denote   the scale factor in the Robertson-Walker (RW) metric as $a(t)\equiv Ae^{\alpha(t)}$, where $\alpha(t)$ is a real function.   By making analytic continuation of  the Lagrangian in the gravitational action to the complex plane of   $\alpha$,  terms corresponding to a total action $I$  may be obtained. The resulting  model \cite{mvjkbj1,mvjkbj2,mvjkbj3} has an earlier  coasting contraction followed by a coasting expansion, with a smooth bounce in between. This model  faces no cosmological problems such as flatness, horizon, etc.  and  is  a viable alternative to the  $\Lambda$-CDM cosmological model in explaining  various observational data, as shown in several works \cite{mvjapj1,mvjapj2,melia1,melia2,melia3,mvjmnras}. We shall here review the  most promising aspects of the resulting cosmological model.

In this paper we also attempt to  perform the quantisation of this  cosmological model using the operator method.   For any cosmological model obtainable from an action principle, the first step is to construct the Hamiltonian  $H$ corresponding to the Lagrangian function $L$. However,  the gravitational  Hamiltonian function  must obey $H=0$ to have the property of time-reparametrisation  invariance for the action. To write a quantum wave equation using this Hamiltonian, the conventional approach is to write

\begin{equation}
H\Psi =0, \label{eq:wd}
\end{equation}
which is the Wheeler-De Witt (WD) equation. In this case, the quantum wave function $\Psi$   corresponds to  a  state with zero energy.  In quantum cosmology, the question then arises  whether this equation applies also to the late universe, which appears to be classical. In the conventional approach, it is  considered that this is possible only if the probability density corresponding to the quantum wave function $\Psi$ is strongly peaked around the trajectories identified by the classical solutions \cite{halli}. In an earlier work, we wrote down a  wave equation for the complex cosmological model  \cite{mvjkbj2} and checked whether it has classical correspondence in the late epochs. This was only partially successful, for we could get only an approximate solution that describes  the late classical epoch  of the universe. In \cite{mvjgrco}, using the `quantum trajectories' approach, we checked whether the coasting evolution of a universe that starts from  singularity \cite{mvjkbj3} will have exact quantum-classical correspondence  and obtained a positive result.  In that case, we have used the de Broglie-Bohm (dBB) \cite{db,bohm} and modified de Broglie-Bohm (MdBB)  \cite{mvjqm1,mvjqm2,mvjqm3,mvjqm4,mvjqm5,mvjqm6} trajectory formulations of quantum mechanics.  In the present paper, instead of writing the WD equation,  we employ the operator method in canonical quantisation and show that the quantum cosmological wave function can be obtained as an exact solution to the corresponding  wave equation for the universe.   Using the MdBB trajectory formulation that envisions complex quantum trajectories, we  show that the complex universe has  exact quantum-classical correspondence throughout its evolution.

The paper is planned as follows. In the next section, we describe the classical field equations for a conventional Friedmann-Lamaitre-Robertson-Walker (FLRW) cosmological model with a scalar field and another one with matter/energy density varying as $a^{-n}$. In Sec. 3, the  action principle that leads to the bouncing and coasting model is discussed. The operator method of canonical quantisation of the model is presented in Sec. 4, along with the solution of the new wave equation for the universe. Sec. 5 is  to demonstrate the exact quantum-classical correspondence for the solution. The last section summarises  the results.    

\section{Cosmological equations}

The Einstein equation in general theory of relativity is obtained by requiring that the action
\cite{landau,weinbook,kolbturner}, 

\begin{equation}
I = \frac {-c^3}{16\pi G}\int   \; R(g_{\mu \nu})\; \sqrt {-g}\; d^{4}x + \frac{1}{c}
 \int\;  \Lambda \; \sqrt {-g} \; d^{4}x
\equiv I_{G}+I_{M} ,   \label{eq:totact}
\end{equation}
be stationary under variation of the dynamical variables in it. The first integral is  the
gravitational action $I_{G}$, where $R(g_{\mu \nu})$ is the curvature
scalar. In the second integral in Eq. (\ref{eq:totact}), which is the matter action
$I_{M}$, $\Lambda $ corresponds to the matter fields  that curve spacetime. 

As an example for $\Lambda$ in cosmology, consider the  case in which there is only a scalar field $\phi (x^{\mu})$  contributing to the energy-momentum of the universe. Here, $\Lambda _{\phi }$ takes the form

\begin{equation}
\Lambda _{\phi } = \frac {\hbar^2}{2m} g^{\mu \nu} \frac {\partial \phi}{\partial
x^{\mu}} \frac {\partial \phi }{\partial x^{\nu}} - V(\phi ),
\label{eq:lambdaphi}
\end{equation}
where $V(\phi )$ is the  density corresponding to the potential energy of the field.   Let  the  distribution of the scalar field $\phi$ be homogeneous and isotropic, such that its density depends only on time $t$. Then we can   restrict ourselves to the RW metric and write down  explicitly the action $I$. Under the Arnowitt-Deser-Misner (ADM) 3+1 split of spacetime \cite{kolbturner},
the curvature scalar   is of the form

\begin{equation}
R=\frac {6}{N^{2}c^2} \frac {\dot {a}^{2}}{a^{2}}- \frac {6k}{a^{2}}.
\label{eq:ricci} 
\end{equation}
 The total action for this case is then

\begin{equation}
I = \int N\sqrt {h}
\left[ -\frac
{c^3}{16\pi G}\left( \frac {6}{N^{2}c^2}\frac {\dot {a}^{2}}{a^{2}} - \frac
{6k}{a^{2}} \right) + \frac{1}{c} \left(\frac{\hbar^2}{2mc^2} \frac {\dot {\phi }^{2}}{N^{2}}-V(\phi
)\right)\right] d^{4}x. \label{eq:iel}
\end{equation}
Here  $N$ is the lapse function, which ensures that the action is time-reparametrisation invariant. It means that using a new time variable $t^{\prime}$ such that $dt^{\prime}=Ndt$ will not affect the equation of motion. Also, $h$ is the determinant of the metric induced on the 3-space, with $\sqrt{-g}=N\sqrt{h}$. Integrating the space part, we get

\begin{eqnarray}
I & = & 2\pi ^{2}\int Na^{3}\left[ -\frac {c^4}{16\pi G}\left( \frac
{6}{N^{2}c^2}\frac {\dot {a}^{2}}{a^{2}} - \frac {6k}{a^{2}} \right) + 
\left(\frac{\hbar^2}{2mc^2} \frac {\dot {\phi
}^{2}}{N^{2}} - V(\phi )\right)\right] dt  \nonumber \\
& \equiv & \int L\; dt. \label{eq:ielphi}
\end{eqnarray}
Varying this action with respect to $N$, $a$ and $\phi$, and keeping the gauge $N=1$, we get the Einstein equations for this case as

\begin{equation}
\frac {\dot {a}^{2}}{a^{2}} + \frac {kc^2}{a^{2}} = 
 \frac {8\pi G}{3c^2}  \left[\frac{\hbar^2}{2mc^2}  {\dot {\phi }^{2}} + V(\phi
)\right] ,\label{eq:t-tphi}
\end{equation}

\begin{equation}
2\frac {\ddot {a}}{a} + \frac {\dot {a}^{2}}{a^{2}} + \frac
{kc^2}{a^{2}} = - \frac{8\pi G}{c^2} 
\left[\frac{\hbar^2}{2mc^2}  {\dot {\phi }^{2}} - V(\phi )\right] , \label{eq:s-sphi}
\end{equation}

and

\begin{equation}
\ddot {\phi }+ 3 \frac {\dot {a}}{a}
\dot {\phi } + \frac{mc^2}{\hbar^2}\frac {dV(\phi ) }{d\phi }=0. \label{eq:consphi}
\end{equation}
In inflationary cosmologies, such a scalar field leads to an exponential expansion during the very early epoch of the universe. One only needs to have an appropriate scalar field potential $V(\phi)$ and suitable initial conditions for inflation to occur in this scenario.

Another example for $\Lambda$ we consider here is that of  a homogeneous and isotropic distribution of matter/energy in the universe, whose density $\rho$ varies as $a^{-n}$. It can be seen to obey an action principle, with action $I$ having the same form as in  (\ref{eq:totact}), with

\begin{equation}
\Lambda =-c^2\frac{C_n}{a^n}, \label{eq:LambdaI}
\end{equation}
where $C_n$ is some constant such that $C_n/a^n$ has dimensions  of density. In this case, the Lagrangian $L$ can be found to be

\begin{equation}
L=L_G+L_M=2\pi^2c^2a^3N\left[ \frac{-1}{16\pi G}\left(\frac{6}{N^2}\frac{\dot{a}^2}{a^2}-\frac{6kc^2}{a^2}\right) -\frac{C_n}{a^n}\right]
\end{equation}
Varying the action with respect to $N$ and $a$ gives

\begin{equation}
\frac{\dot{a}^2}{a^2}+\frac{kc^2}{a^2}=\frac{8\pi G}{3}\frac{C_n}{a^n}
\end{equation}
and 

\begin{equation}
2\frac{\ddot{a}}{a}+\frac{\dot{a}^2}{a^2}+\frac{kc^2}{a^2}=-\frac{8\pi G(n-3)}{3}\frac{C_n}{a^n}
\end{equation}
These equations agree with the field equations for a perfect fluid if the  pressure of the fluid  is

\begin{equation}
p=\frac{(n-3)}{3}c^2\frac{C_n}{a^n}
\end{equation}
so that the equation of state parameter, defined by $p=w\rho c^2$ is $w=(n-3)/3$.

It is interesting to note that  in the case of RW metric, if we introduce a new variable $\alpha$ by denoting $a(t)\equiv Ae^{\alpha(t)}$ with $A$ as some constant,   the gravitational action $I_G$  can be cast in the form of $I_M$ \cite{halli}. In this case, one can write

\begin{equation}
I_G= -\frac{3\pi}{4G}  \int Nc^4{ a}^3 \left(\frac{\dot{\alpha}^2}{N^2c^2}-\frac{k}{A^2e^{2\alpha}}\right)dt
\end{equation} 
Here the action $I_G$ looks like $I_M$ in (\ref{eq:ielphi}) for a scalar field $\alpha $ with  potential $V(\alpha)= \frac{k}{A^2e^{2\alpha}}$.  Considering the fact that $I_M$ in this case is capable of giving the quantum wave equation to be satisfied by the scalar field, one can take this as suggestive of some close relation between general relativity and quantum mechanics.

In the next section, we shall make use of this important observation to describe the evolution of the universe with an action principle that contains only a single term $I_G$ in $I$.

\section{A new scalar field}

 As in the above section, we write the scale factor  as $a(t)=A e^{\alpha(t)}$ and let $a(0)=A$. In the previous work \cite{mvjkbj1,mvjkbj2}, we introduced a new field $\beta$ by considering the analytic continuation of the parameter $\alpha$ to the complex plane to get  $\phi(t)=\alpha(t)+i\beta(t)$ and assumed that the total action $I$ consists of only the part $I_G$.   Thus   the line element becomes

\begin{equation}
ds^{2} = c^2dt^{2} - a^{2}(t)e^{2i\beta(t)} \left[ \frac {dr^{2}}{1-kr^{2}} +
r^{2} (d\theta ^{2} + \sin ^{2}\theta\; d\phi ^{2})\right],
\label{eq:rwle_beta} 
\end{equation} 
 We may now denote  ${\tilde a}(t)\equiv a(t)e^{i\beta(t)} = A e^{\phi(t)}$ and write the  action  as
 
 \begin{eqnarray}
I& = &I_G= \int N\sqrt {h}
\left[ -\frac
{c^3}{16\pi G}\left( \frac {6}{N^{2}c^2}\frac {\dot {\tilde a}^{2}}{{\tilde a}^{2}} - \frac
{6k}{{\tilde a}^{2}} \right) \right] d^{4}x \nonumber \\
 & = & 2\pi ^{2}\int N{\tilde a}^{3}\left[ -\frac {c^4}{16\pi G}\left( \frac
{6}{N^{2}c^2}\frac {\dot {\tilde a}^{2}}{{\tilde a}^{2}} - \frac {6k}{{\tilde a}^{2}} \right) \right] dt    \label{eq:ig} 
\end{eqnarray}
 From now onwards, let us consider only a closed universe. If we vary $I_G$    with respect to $N$ and $\tilde{a}$, with $k=+1$, the Einstein field equation gives
 
 \begin{equation}
\left( \frac { \dot {\tilde {a}}}{\tilde {a}} \right) ^{2} + 
\frac {c^2}{\tilde {a}^{2}} = 0 \label{eq:t-tc}
\end{equation}
and

\begin{equation}
2\frac {\ddot {\tilde {a}}}{\tilde {a}} + \left( 
\frac {\dot {\tilde {a}}}{\tilde {a}}\right) ^{2} + \frac {c^2}{\tilde
{a}^{2}} = 0. \label{eq:s-sc} 
\end{equation}
 Using ${\tilde a}\equiv Ae^{i \phi}$, one can equivalently write this action in (\ref{eq:ig})  with   $\phi$ as the only field variable. We thus write the same action,  with $k=+1$, as

\begin{equation}
I=I_G= -\frac{3\pi}{4G}  \int Nc^4{\tilde a}^3 \left(\frac{\dot{\phi}^2}{N^2c^2}-\frac{1}{A^2e^{2\phi}}\right)dt
\end{equation}
 Varying this with respect to $N$ and $\phi$, one gets the field equations

\begin{equation}
\dot{\phi}^2+\frac{c^2}{A^2e^{2\phi}}=0 \label{eq:t-tc_phi}
\end{equation}
and 

\begin{equation}
\ddot{\phi}+\frac{3}{2}\dot{\phi}^2+\frac{c^2}{2A^2e^{2\phi}}=0 \label{eq:s-sc_phi}
\end{equation}
These equations  are equivalent to equations (\ref{eq:t-tc}) and (\ref{eq:s-sc}), respectively. We also find  that the last equation resembles  (\ref{eq:consphi}), which is the  conservation law for the scalar field.

 It can be seen that  the above field equations  have  solutions
 
 \begin{equation}
 {\tilde a}\equiv A e^{\phi}=a_0\pm ict, \qquad \hbox {and} \qquad \phi=\log\left(\frac{a_0\pm ict}{A}\right), \label{eq:soln_a}
 \end{equation}
respectively,  where $a_0$ is some real constant. These two solutions are equivalent and it is  easy to see that  $\; a_0=A$. The scale factor $a(t)$ of the  model is now 
 
 \begin{equation}
 a=\sqrt{a_0^2+c^2t^2}.
 \end{equation}
 This corresponds to a bouncing nonsingular universe, with minimum for $a(t)$ as  $a_0$, at $t=0$. (See Figure 1.)   For $ct \gg a_0$, the model coasts ($a\propto t$) eternally.  A linear evolution of the scale factor is characteristic of the Milne model, but the present one cannot be compared with it, for the Milne model does not take account of gravity as described in general relativity.
 
 \begin{figure}[ht] 
\centering{\resizebox {1 \textwidth} {0.4 \textheight }  
{\includegraphics {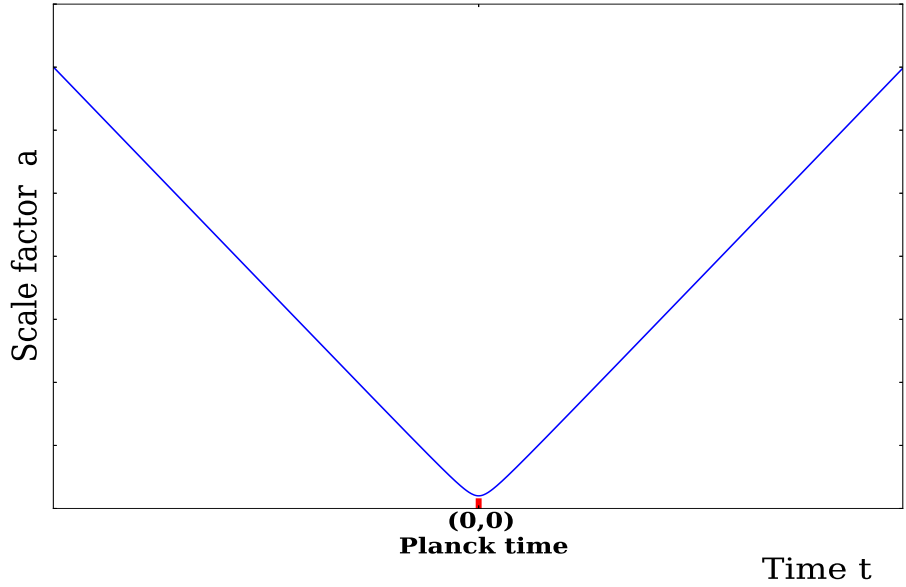}} 
\caption{The variation of scale factor with time. The smooth bounce occurs in between the coasting contraction  and the coasting expansion phases. 
} }  \label{fig:bounce}
  \end{figure} 
 
 Surprisingly,  the field equations in the above case can be rewritten in a form similar to that of   a FLRW model.  When redefined to have appropriate dimensions, the variable $\beta$ would correspond to a scalar field that fills this universe.  By separating the real and imaginary parts of  them, we obtain

\begin{equation}
\frac {\dot {a}^{2}}{a^{2}} + \frac {c^2}{a^{2}} = \dot {\beta }^{2} +
\frac {2c^2}{a^{2}} \sin ^{2}\beta , \label{eq:t-tbeta}
\end{equation}

\begin{equation}
2\frac {\ddot {a}}{a}+\frac {\dot {a}^{2}}{a^{2}} + \frac {c^2}{a^{2}}
= 3\left( \dot {\beta }^{2} + 
\frac {2c^2}{3a^{2}} \sin ^{2}\beta \right) , \label{eq:s-sbeta}
\end{equation}

\begin{equation}
\ddot {\beta } + 2\dot {\beta }\frac {\dot {a }}{a} =0,
\label{eq:consbeta} 
\end{equation}
and

\begin{equation}
2\dot {\beta }\frac {\dot {a}}{a} = \frac {c^2}{a^{2}}\sin 2\beta
.\label{eq:betadot} 
\end{equation}
These equations are similar to equations (\ref{eq:t-tphi}) - (\ref{eq:consphi}) for a universe filled with a scalar field.   We may get the solution to $\beta (t)$ from the above equations as

\begin{equation}
\dot {\beta }(t) = \frac {\pm c\; a_{0}}{a^{2}(t)} = 
\frac {\pm c \; \cos ^{2} \beta }{a_{0}},
\end{equation}
and

\begin{equation}
\beta (t) = \tan ^{-1} ( \frac {\pm ct} {a_{0}}),
\end{equation}
Substituting these solutions for $\beta$ and $\dot{\beta}$ into equations (\ref{eq:t-tbeta}) and (\ref{eq:s-sbeta}), one can  write

\begin{equation}
\frac{{\dot a}^{2}}{a^{2}}  + \frac {c^2}{a ^{2}} = \frac {2c^2}{a^{2}} - \frac 
{c^2 a^{2}_{0}}{a^{4}}  \label{eq:constrainte}
\end{equation}

\begin{equation}
2\frac{\ddot a}{a}+\frac {\dot {a}^{2}}{a^{2}}  + \frac {c^2}{a ^{2}} =
\frac {2c^2}{a^{2}} + \frac  
{c^2 a^{2}_{0}}{a^{4}}, \label{eq:fe}
\end{equation}
We now note that these equations  result also from the variation of an action  in  (\ref{eq:totact}),  with the form of $\Lambda$ slightly modified than that given in (\ref{eq:LambdaI}). Here we must use

\begin{equation}
\Lambda =c^2\left( \frac{C_4}{a^4} -\frac{C_2}{a^2}\right),
\end{equation}
where $C_2=3c^2/4\pi G$ and $C_4=3c^2a_0^2/8\pi G$. Such a real action that leads to the same equations (\ref{eq:constrainte})-(\ref{eq:fe})  corresponds to a closed ($k=+1$) FLRW model with  total
energy density and pressure  given by 

\begin{equation}
{\rho}  = \frac{3c^2}{8\pi G}\left( \frac{2}{a^{2}} -
\frac{ a^{2}_{0}}{a^{4}}\right) ,\label{eq:rhot}
\end{equation}

\begin{equation}
{p}   ={-\frac{c^4}{8\pi G}}\left( \frac {2}{a^{2}} + 
\frac{a^{2}_{0}}{a^{4}} \right) \label{eq:pt}
\end{equation}
respectively.  This physical model with the same evolution $a(t)=\sqrt{a_0^2+c^2t^2}$ can be shown to be a viable one with several positive features when compared to other models. For $ct\gg a_0$, this model coincides with an eternal coasting cosmological model discussed in \cite{mvjkbj3}, for the special case of $k=+1$. It may also be noted that another special case of the eternal coasting model in \cite{mvjkbj3} (one with $k=0$) is examined in detail under the title of `$R_h=ct$ cosmological model' \cite{melia1,melia2,melia3}.

The quantities in equations (\ref{eq:rhot}) and (\ref{eq:pt})   correspond to a conserved total energy-momentum tensor. However, we may consider the energy corresponding to  this to be made up of constituents, that may or may not be separately conserved. An obvious conserved constituent of the above energy density is a negative energy density $\rho_{-}$, given by

\begin{equation}
\rho_{-}= -\frac{3c^2}{8\pi G} \frac{ a^{2}_{0}}{a^{4}}.
\end{equation}
This has pressure, given by

\begin{equation}
p_{-}= -\frac{c^4}{8\pi G} \frac{ a^{2}_{0}}{a^{4}}
\end{equation}
so that $p_{-}=(1/3)\rho_{-}c^2$. This negative energy, which closely resembles that of the Casimir energy \cite{casimir}, has its density   significant (when compared with the remaining part) only near the bounce \cite{mvjthesis}. Such a bounce  is sometimes referred to as a `Tolman wormhole' \cite{coule}. Another  feature of the present model  in connection with this energy density is that the gravitational charge ${\rho} c^2+3{p}$ for the model is

\begin{equation}
{\rho} c^2+3{p}=-\frac{3c^4}{4\pi G}\frac{{a_0}^2}{a^4}.
\end{equation} 
Since the negative energy density disappears very quickly for $ct\gg a_0$,   also the gravitational charge vanishes for  $a(t)\gg a_0$. This leads to the coasting evolution for the very early/late epochs.

 It was pointed out in \cite{mvjkbj1,mvjkbj2,mvjkbj3,mvjmnras} that the remaining (conserved) part of the total density in (\ref{eq:rhot}) (that varies  as $ a^{-2}$) can be considered to be made up of  ordinary matter (with equation of state $p_m=0$ or $p_m=\rho_m/3$) and a repulsive dark energy (with equation of state $p_{\Lambda}=-\rho_{\Lambda}c^2$). This can happen if there is creation of matter at the expense of the (decaying) dark energy. In the present epoch of the universe, the rate of creation of matter need only to be very small, so that it will  hardly be  observable. 

Cosmological observations suggest that   at the present epoch,  densities of matter and dark energy have nearly the same magnitude. If this feature is true only for the present universe,  such  a near equality at the present epoch owes an explanation and this is referred to as the `coincidence problem'. It is easy to see that unless the dark energy density and the matter density vary in the same manner, there will appear a coincidence problem.  We see that in the present model, no coincidence problem appears since one has both  components varying as $a^{-2}$.

 It may also be noted that  this model is devoid of  the so-called synchronicity problem \cite{kirshner}. This  is related to the value of the combination of physical quantities $H_0$, the present value of the Hubble constant and $t_0$, the  age of the universe, estimated using direct methods. The  observed value of the combination $H_0t_0$ is very close to unity, which is another coincidence since it could not have occurred at any other epoch in the universe. In the present model, this is however not a coincidence; this is  true for all times and is a prediction of the model. Here it must be true at all epochs except during the initial bounce.

 In addition to the absence of coincidence and synchronicity problems, the model is devoid of all problems discussed in Friedmann cosmology in the 1980s and 90s, such as the flatness, horizon, size and monopole problems. The present cosmological model  agrees  well with several cosmological observations, as discussed extensively in the literature \cite{mvjapj1,mvjapj2,melia1,melia2,melia3,mvjmnras}.

\section{Quantisation of the model}
 In order to quantise this model, let us start from the Lagrangian in $I_G$, the gravitational action, for a universe described by RW metric with real scale factor $a$.
  This Lagrangian   may be written as
 
 \begin{equation}
 L_G=\frac{3\pi}{4G}c^4 N \; { {a}^{3}} \left[ 
\frac {1}{N^{2} c^2} \left( \frac { \dot { {a}}}{ {a}}\right)^{2} - \frac
{1}{{ {a}^{2}}} \right], \label{eq:L_Gquantum}
 \end{equation}
omitting the negative sign. We note that $\dot{N}$ does not appear in this expression and hence the momentum conjugate to $N$, denoted by $\pi_N$, vanishes.  The  conjugate momentum to $  {a} $ is

\begin{equation}
\pi _{  {a}} = \frac { \partial L_G }{ \partial \dot { {a}}} 
=\frac {3 \pi}{2G} \frac{c^2}{N} {a} \dot { {a}}. \label{eq:pi}
\end{equation}
The  canonical Hamiltonian is then

\begin{equation}
{ H_c} =N\left( 
\frac {3\pi c^4}{4G}   {a}+\frac {G}{3 \pi c^2} \frac {\pi _{ {a}}^{2}}{ {a}}\right) \equiv NH. \label{eq:can_H}
\end{equation} 
Since $\pi_N$ vanishes for all times, its time derivative, given by the Poisson bracket, also vanishes. That is,

\begin{equation}
\dot{\pi}_N=\{H_c,\pi_N\}=\frac{\delta H_c}{\delta N}=0.
\end{equation}
This leads to $H=0$, where $H$ is defined by (\ref{eq:can_H}). Such a classical constraint equation can be seen to be the consequence of  time-reparametrisation invariance. In other words, $H=0$  ensures that using  $dt^{\prime}=Ndt$ will not affect the equation of motion. Also, this allows us to choose some convenient gauge for $N$. Note that we have already chosen $N=1$ in the above sections. However,  in the following discussion, we do not intend to set  any such gauge for $N$.

 The quantum theory of this model can now be formulated with ${a}$ as the field variable. The first step in quantising the classical model is to consider all observables, such as ${ a}$, $\pi_{ a}$, etc., as operators. The usual approach is to write the   Wheeler-De Witt (WD) equation, given in Eq. (\ref{eq:wd}),
 to take account of the classical constraint relation $H=0$. This is considered to be the wave equation for the universe,  defined in a minisuperspace with only one coordinate ${a}$.  However, here an operator-ordering ambiguity arises  while replacing    $\pi_{ a}^2/{ a}$ with differential operators. The usual approach is to adopt a factor-ordering  parametrised by a constant $r$ \cite{kolbturner}, such that 
 
 \begin{equation}
 \pi_{ a}^2\rightarrow -\hbar^2 {a}^{-r}\left(\frac{\partial}{\partial { a}}{ a}^r\frac{\partial}{\partial { a}}\right). \label{eq:op_order}
 \end{equation}
Various choices for the value of $r$ are  made, hoping that its value will not significantly affect  semiclassical calculations. Instead of pursuing this approach, in the present work, we  adopt the well-known operator method to address the quantisation of the model. Here, one assumes  ${ a}$ and $\pi_{ a}$ as operators that obey the fundamental commutation relation 
 
\begin{equation}
[{ a},\pi_{ a}]=i\hbar . \label{eq:comm1}
\end{equation} 
In addition, let us postulate a commutation and an anticommutation relation between ${\pi_{ a}^2}/{{a}}$ and ${ a}$ as

\begin{equation}
\left[\frac{\pi_{ a}^2}{{a}}, { a}\right]=0. \label{eq:comm2}
\end{equation}
and 

\begin{equation}
\left\lbrace\frac{\pi_{ a}^2}{{a}}, { a}\right\rbrace=2 \pi_{ a}^2, \label{eq:pi_sq}
\end{equation}
respectively.  This  is  useful in avoiding any operator-ordering ambiguity and helps circumventing  arbitrary steps as in (\ref{eq:op_order}).     Using Eq. (\ref{eq:comm2}), one can see that also ${ a}$ and $H$ are commuting observables. i.e.,
 
 \begin{equation}
 [{{a}},H]=0. \label{eq:comm_aH}
 \end{equation}
The result that ${ a}$ and $H$ commute does not lead to any conceptual problems; it only endorses the fact that the universe can have simultaneous  eigenstates of ${a}$ and $H$. With this additional input, we  define two new operators,

\begin{equation}
C=\frac{1}{\sqrt{\hbar c}}\left( \sqrt{\frac{3\pi c^4}{4G}}{a}+i\sqrt{\frac{G}{3\pi c^2}}\pi_{ a}\right),
\end{equation}
and 

\begin{equation}
C^{\dagger}=\frac{1}{\sqrt{\hbar c}}\left( \sqrt{\frac{3\pi c^4}{4G}}{a}-i\sqrt{\frac{G}{3\pi c^2}}\pi_{ a}\right).
\end{equation}
We may now obtain the relation

\begin{equation}
C^{\dagger}C=\frac{1}{\hbar c}\left(\frac{3\pi c^4}{4G}{ a}^2+\frac{G}{3\pi c^2}\pi_{ a}^2-\frac{\hbar c}{2}\right).
\end{equation} 
Using the definition (\ref{eq:can_H}) for $H$ and the relations (\ref{eq:comm2}) and (\ref{eq:pi_sq}), one can write this expression as

\begin{equation}
C^{\dagger}C=\frac{1}{\hbar c}\left({a} H-\frac{\hbar c}{2}\right).
\end{equation}
 It may also be noted that 

\begin{equation}
[{a},C^{\dagger}C]=0, \qquad [H,C^{\dagger}C]=0, \qquad  [C,C^{\dagger}]=1, \label{eq:comm_other}
\end{equation}
and

\begin{equation}
{ a}H={\hbar c}\left(C^{\dagger}C +\frac{1}{2}\right).
\end{equation}
$C^{\dagger}C$   can now be considered as a number operator, with 

\begin{equation}
C^{\dagger}C \ket{n} =n\ket{n}  ,
\end{equation}
so that 

\begin{equation}
{ a}H\ket{n}={\hbar c}\left(n+\frac{1}{2}\right)\ket{n}. \label{eq:eigenval_eqn}
\end{equation}
With the help of the commutation  relations (\ref{eq:comm1}),  (\ref{eq:comm2}), (\ref{eq:pi_sq}),  (\ref{eq:comm_aH}) and (\ref{eq:comm_other}), one can write

\begin{equation}
{ a}H(C\ket{n})={\hbar c}\left[(n-1)+\frac{1}{2}\right](C\ket{n}),
\end{equation}
showing that $C$ is a lowering operator. Similarly,

\begin{equation}
{ a}H(C^{\dagger}\ket{n})={\hbar c}\left[(n+1)+\frac{1}{2}\right](C^{\dagger}\ket{n}).
\end{equation}
showing that $C^{\dagger}$ is a raising operator. The ground state of the universe, with $n=0$,  may then obey

\begin{equation}
C\ket{0}=0.
\end{equation}
 We can now write a differential equation that corresponds to this equation, without any operator-ordering ambiguity. This is   as in the case of harmonic oscillator, but in terms of the minisuperspace coordinate ${a}$. The  resulting differential equation, with $\braket{{ a} |{0}}\equiv \Psi_0({ a})$, is

\begin{equation}
\frac{d\Psi_0({ a})}{d{ a}}+\frac{3\pi c^3}{2G\hbar} \;{a}\; \Psi_0({a})=0,
\end{equation}
 whose solution is
 
 \begin{equation}
 \Psi_0({ a})=\exp\left(-\frac{{ a}^2}{2\lambda^2}\right), \label{eq:soln}
 \end{equation}
 where $ \lambda^2\equiv \frac{2G\hbar}{3\pi c^3}$.  This solution is similar to that of the ground state harmonic oscillator. One can also see that this  is the solution for  an eigenvalue equation, 
 
 \begin{equation}
{a} H\ket{0}=\frac{\hbar c}{2}\ket{0},
 \end{equation}
 with   ${\hbar c}/{2}$ as the eigenvalue. Then we can write the wave equation for the universe in the form
 
 \begin{equation}
{a} H\Psi_0({a})=\frac{\hbar c}{2}\Psi_0({a}). \label{eq:ground}
 \end{equation}
 Note that this equation is different from the WD equation (\ref{eq:wd}), which is based on $H=0$.    In fact, the zero energy condition $H=0$  is classical and one need not expect this to be exactly satisfied in a quantum theory too. In addition, the above equation does not contain time as a parameter and hence it is time-reparametrisation invariant, just as the WD equation. We claim that (\ref{eq:ground}) is a more appropriate wave equation for the universe than the latter. 
 
  It may  be noted that (\ref{eq:ground}) rectifies  a similar equation used  in \cite{mvjkbj1,mvjkbj2}. In this reference, we anticipated the presence of a  zero-point energy $\epsilon_0$, in a somewhat  arbitrary manner. But in that case, we could not get an exact solution for the universe and hence no exact classical-quantum correspondence.   In this paper we have shown, using the operator method, that the quantum ground state of the present model can be described by an exact solution of the problem, which is fit for the universe.

 In the next section, we shall  see  that the results obtained above for the complex universe, based on the  operator method, has exact classical-quantum correspondence.

 \section{Classical-quantum correspondence}
 
 We have obtained the ground ($n=0$) state of the universe as  described by a Gaussian wave function, and now its correspondence  with the classical solution needs to be verified. One  expects that at least for large values of $a$, the wave function would correspond with the classical world. The usual approach to this  is to check whether the probability distribution obtained from the wave function exactly peaks over a classical trajectory. However, it may be noted that here  we have only one state for the universe and even the definition of probability in this context is quite vague. An alternative approach  is to obtain the quantum trajectories of the given wave function and to check whether the trajectories agree with the classical ones, at least in the limiting case. We have seen in \cite{mvjgrco} that the coasting evolution of the universe that starts from a singularity will have exact classical-quantum correspondence when checked in this manner.   We show, using the MdBB complex trajectory formulation of quantum mechanics, that in the present case of complex  cosmology too, there is exact correspondence between the quantum and classical description of the universe, throughout its history of evolution.
 
 To demonstrate this, we make use of  the equation of motion used in the  trajectory  approaches, given by
 
 \begin{equation}
 \pi_{{a}}=\frac{\partial S}{\partial {a}}, \label{eq:pi2}
 \end{equation}
where $S$ is the action.  In the MdBB approach, where the wave function  is identified to be of the form $\Psi ({ a}) =e^{iS({ a})/\hbar}$ \cite{mvjqm1}, the action $S$ may be found   to be 
 
 \begin{equation}
 S=i\hbar  \frac{{a}^2}{2{\lambda}^2}.
 \end{equation}
Since in general $S$ is a complex function, the momentum computed using (\ref{eq:pi2}) and also the trajectories may be complex.  In the present case, the trajectories are obtainable from 
 
\begin{equation}
\pi_a=\frac{\partial L_G}{\partial a}=\frac{3\pi}{2G}c^2a\dot{a}
\end{equation} 
Henceforth, we shall denote the complex trajectories as $\tilde{a}(t)$. Using (\ref{eq:pi2}), one  obtains
 
 \begin{equation}
 \dot{\tilde{a}}=icN.
 \end{equation}
 On integrating, we get
 
 \begin{equation}
 \tilde{a}=a_0+icNt.
 \end{equation}
 Retaining the original  negative sign in Eq. (\ref{eq:L_Gquantum}) (instead of omitting it) would be equivalent to writing $-N$ in place of $N$ in $L_G$. In that case, we get the above result as $\tilde{a}=a_0-icNt$. In the gauge $N=+1$, this is the same as the solution for ${\tilde a}$ in (\ref{eq:soln_a}),  which is the desired result. Thus the  trajectories corresponding to the ground state of the  quantum universe is precisely the same as that in the classical case and we have exact quantum-classical correspondence.

\section{Discussion}

Einstein considered the left hand side of his equation as nice and geometrical, while according to him the right hand side, which contains the energy-momentum tensor, is  somewhat less compelling \cite{seancarroll}.  He is said to have even fancied that also  $T_{\mu \nu}$  somehow followed  from geometry.  To our knowledge, there were no such attempts to address this possibility even for specific instances, other than that  in the previous work \cite{mvjkbj1,mvjkbj2}.  Here we showed that formulating a theory of gravity based solely  on the gravitational action $I_G$, which contains only geometric terms, is possible at least in the case of cosmology. Analytic continuation of the Lagrangian function in $I_G$ to the complex plane of $\alpha(t)$, where the scale factor $a(t)=Ae^{\alpha(t)}$ in the $k=+1$ RW metric,  results  in  the same field equations as  in a closed FLRW universe  with  homogeneous and isotropic distribution of a new scalar field $\beta (t)$. Solving the Einstein equation in this case gives a bouncing and eternal coasting cosmological model. This model is shown to follow also from the   action for a closed universe with homogeneous and isotropic distribution of energy/matter, with density varying as $a^{-2}$ and an additional negative energy density that varies as $a^{-4}$. The latter energy density is significant only near the bounce. This  was shown to be a viable alternative to the $\Lambda$-CDM model in explaining the evolution of the background spacetime, while not exhibiting any of the conceptual problems that plague the $\Lambda$-CDM model. We developed a quantum theory for the complex model using an approach different from that of the Wheeler-DeWitt equation.  While resorting to the operator method of canonical quantisation based on the fundamental commutation relations, it helped to circumvent the operator ordering ambiguity in the WD equation. Using this approach, we obtained a quantum wave equation for the universe, which is a modified WD equation. We find that the exact solution of this equation is an acceptable wave function for the universe. Most notably, we have shown that the solution of the present quantum wave equation  corresponds to  the classical solution in all respects, as per the MdBB trajectory formulation of quantum mechanics.  

\section{Acknowledgements}
{I wish to thank Professors K. Babu Joseph and K. P. Satheesh for enlightening discussions.}

\section{Declarations}

{\bf Funding} There was no funding for this work from any source.

{\bf Conflicts of interest/Competing interests} The author declares no conflict of interests.

{\bf Availability of data and material } The work does not use any data or material directly.

{\bf   Code availability} The work does not rely on any codes.

\section{References}


\end{document}